\documentclass[aps,prb,superscriptaddress,showkeys]{revtex4-2}
\usepackage[utf8]{inputenc}
\usepackage{amsmath,amssymb,amsfonts}
\usepackage{multirow}
\usepackage{graphicx,float}
\usepackage[colorlinks=true,
citecolor=blue,
linkcolor=blue,
urlcolor=blue]{hyperref}
\usepackage{footmisc}
\usepackage{array}
\newcolumntype{P}[1]{>{\centering\arraybackslash}p{#1}}

\begin{document}
	
\title{Understanding the Role of Four-Phonon Scattering in the Lattice Thermal Transport of Monolayer MoS$_{2}$}
\author{Saumen Chaudhuri}
\affiliation{Department of Physics, Indian Institute of Technology Kharagpur, Kharagpur 721302, India}
\author{Amrita Bhattacharya}
\email[corresponding author: ]{b\_amrita@iitb.ac.in}
\affiliation{Department of Metallurgical Engineering and Materials Science, Indian Institute of Technology Bombay, Mumbai 400076, India}
\author{{A. K. Das}}
\affiliation{Department of Physics, Indian Institute of Technology Kharagpur, Kharagpur 721302, India}
\author{{G. P. Das}}
\email[corresponding author: ]{gourpdas@gmail.com}
\affiliation{Research Institute for Sustainable Energy, TCG Centres for Research and Education in Science and Technology, Sector V, Salt Lake, Kolkata 700091, India}
\author{B. N. Dev}
\email[corresponding author: ]{bhupen.dev@gmail.com}
\affiliation{Department of Physics and School of Nano Science and Technology, Indian Institute of Technology Kharagpur, Kharagpur, 721302, India}
\affiliation{Centre for Quantum Engineering, Research and Education, TCG Centres for Research and Education in Science and Technology, Sector V, Salt Lake, Kolkata 700091, India}
	
\begin{abstract}
In the calculations of lattice thermal conductivity ($\kappa_{\text{L}}$), vital contributions stemming from four-phonon scattering is often neglected. The significance of four-phonon scattering in the thermal transport properties of monolayer (ML)
MoS$_{2}$ has been unraveled using first-principles calculations combined with the Boltzmann transport equation. If only three-phonon scattering processes are considered then the $\kappa_{\text{L}}$ is found to be significantly overestimated ($\sim$ 115.8 Wm$^{-1}$K$^{-1}$ at 300 K). With the incorporation of the four-phonon scattering processes, the $\kappa_{\text{L}}$ reduces to 24.6 Wm$^{-1}$K$^{-1}$, which is found to be closer to the experimentally measured $\kappa_{\text{L}}$ of 34.5 Wm$^{-1}$K$^{-1}$. Four-phonon scattering significantly impacts on the carrier lifetime ($\tau$) of the low-energy out-of-plane acoustic mode (ZA) phonons and thereby, suppresses its contribution in $\kappa_{\text{L}}$ from 64\% (for three-phonon scattering) to 16\% (for both three- and four-phonon scatterings). The unusually high four-phonon scattering rate ($\tau_{4}^{-1}$) of the ZA phonons is found to result from the simultaneous effect of the acoustic-optical frequency gap, strong anharmonicity and the reflection symmetry imposed selection rule.  The strong coupling between the quadratic dispersion of the ZA mode and the $\tau_{4}^{-1}$ is discovered by the application of  mechanical strain. The strain induced increase in the linearity of the ZA mode dispersion dramatically reduces the significance of the four-phonon scattering in the strained ML-MoS$_{2}$, both qualitatively and quantitatively. These conclusions will provide significant insights to the thermal transport phenomena in ML-MoS$_{2}$, as well as any other 2D material.    
\end{abstract}
\keywords{DFT, four-phonon, MoS$_{2}$, thermal conductivity, mechanical strain}

\date{\today}
\maketitle
	
\section{Introduction}
Monolayer transition metal di-chalcogenides (TMDCs), in the last decade, has attracted immense interest for various practical applications due to its unique physical, chemical and thermal properties \cite{radisavljevic2011integrated, wang2012electronics, johari2011tunable}. Apart from the unique electronic properties, due to the low-dimensional structure and the resulting interface phonon scattering and phonon confinement effects, the two-dimensional (2D) TMDCs comes naturally with a low phonon thermal conductivity ($\kappa_{\text{L}}$) [as shown in \cite{rai2020electronic} and references therein]. In the semiconducting TMDCs with a sizeable band gap, such as MoS$_{2}$, WS$_{2}$, HfS$_{2}$, ZrS$_{2}$ etc., the thermal transport mechanism is almost entirely governed by the lattice contribution ($\kappa_{\text{L}}$) to it, and the electronic contribution ($\kappa_{\text{e}}$) is found to be negligible \cite{chaudhuri2023strain, guo2016biaxial, chaudhuri2023hydrostatic}. Owing to the inverse relationship between $\kappa$ (= $\kappa_{\text{L}}$ + $\kappa_{\text{e}}$) and thermoelectric efficiency, various 2D TMDCs have emerged as a potential thermoelectric material with considerably high efficiency \cite{chaudhuri2023strain, guo2016biaxial, wang2021improved, bera2019strain, chaudhuri2023hydrostatic}. Alongside experiments, first-principle calculations have played a significant role in predicting $\kappa_{\text{L}}$ and thereby, filtering out the best thermoelectric materials \cite{chaudhuri2023strain, guo2016biaxial, wang2021improved, bera2019strain, chaudhuri2023hydrostatic}. However, the accurate prediction of $\kappa_{\text{L}}$ is extremely important, as an under- or over-prediction may lead to misleading conclusions regarding the thermoelectric performance. 
	
Among the various monolayer (ML) TMDCs, the thermal transport properties of MoS$_{2}$ has been explored extensively over the years \cite{chaudhuri2023strain, bhattacharyya2014effect, huang2014theoretical, xiang2019monolayer, chaudhuri2023hydrostatic}. However, there have been no consensus over the value of $\kappa_{\text{L}}$ obtained using various theoretical approaches and from experiments. Mingo \textit{et al.} using first-principle calculations combined with the Boltzmann transport equation (BTE) have reported a $\kappa_{\text{L}}$ of $83$ Wm$^{-1}$K$^{-1}$ for ML-MoS$_{2}$ at $300$ K with a sample size of 10 $\mu$m \cite{li2013thermal}. The value of $\kappa_{\text{L}}$ can be further increased by 30\% by increasing the sample size. There are multiple theoretical articles based on the first-principle BTE calculations reporting a large $\kappa_{\text{L}}$ value of ML-MoS$_{2}$, such as: 116.8 Wm$^{-1}$K$^{-1}$ \cite{sharma2019electron}, 103 Wm$^{-1}$K$^{-1}$ \cite{gu2014phonon} and 131 Wm$^{-1}$K$^{-1}$ \cite{gandi2016thermal}. In fact, the in-plane value of $\kappa_{\text{L}}$ for bulk-MoS$_{2}$ is also found to be as high as 98.9 Wm$^{-1}$K$^{-1}$ \cite{gandi2016thermal}. However, Xing \textit{et al.} measured the thermal conductivity of free-standing single-layer MoS$_{2}$ to be only 34.5 Wm$^{-1}$K$^{-1}$ using temperature dependent Raman spectroscopy \cite{yan2014thermal}. Also, Sahoo \textit{et al.} reported a value of 52 Wm$^{-1}$K$^{-1}$ measured on a substrate-supported few-layer MoS$_{2}$ film \cite{sahoo2013temperature}. Compared to the first-principle based calculations on ML-MoS$_{2}$, the molecular dynamics (MD) simulations are seen to produce results that are in better agreement with experiments, such as: 38.1 Wm$^{-1}$K$^{-1}$ \cite{krishnamoorthy2019thermal}, 26.2 Wm$^{-1}$K$^{-1}$ \cite{wei2014phonon} and 19.76 Wm$^{-1}$K$^{-1}$ \cite{ding2015plane}. Besides, Zhang \textit{et al.} obtained a value of 23.2 Wm$^{-1}$K$^{-1}$ at room temperature using the combination of density functional perturbation theory (DFPT) and non-equilibrium Green’s functions (NEGF) approach \cite{cai2014lattice}. The classical MD simulation results, although ignores the quantum effects in specific heat and phonon scattering, can be considered as a reliable benchmark at 300 K since the Debye temperature of ML-MoS$_{2}$ is only $\approx$ 262 K \cite{peng2016thermal}. Thus, the large discrepancy between the first-principle BTE results and experimentally measured $\kappa_{\text{L}}$ poses a serious question towards the first-principle calculations. Apart from the quantitative disagreement, the description of the role of individual phonon modes in the thermal transport mechanism of ML-MoS$_{2}$ is also baffling. Therefore, it is necessary to investigate the thermal transport properties of ML-MoS$_{2}$ in greater detail and with better accuracy through thorough first-principle calculations. 
	
Over the years, thermal conductivity ($\kappa_{\text{L}}$) calculations based on density functional theory (DFT) have garnered significant popularity due to two reasons: less computational expense and incredible agreement with experimental $\kappa_{\text{L}}$ for variety of systems. To mitigate the computational burden, these calculations consider only the lowest-order perturbative intrinsic scattering involving three phonons. Despite neglecting all the higher-order perturbative terms of the crystal Hamiltonian, the first-principle calculations have proved to be reasonably accurate in predicting the $\kappa_{\text{L}}$ of different materials \cite{broido2007intrinsic, garg2011role, li2012thermal, li2012thermal1}. However, the methodology has become questionable, since first-principle calculations have overestimated the thermal conductivity of a number of materials \cite{feng2016quantum, feng2017four, feng2018four, yang2019stronger, zhang2022four} including ML-MoS$_{2}$ \cite{sharma2019electron, gu2014phonon, gandi2016thermal}. The subsequent incorporation of the next higher-order anharmonic term, which corresponds to scattering processes involving four phonons, significantly reduces the error in the estimation of $\kappa_{\text{L}}$ for diamond \cite{feng2016quantum}, zinc-blende BAs \cite{feng2017four}, AlSb \cite{yang2019stronger} and ML-TaS$_{2}$ \cite{zhang2022four}. For graphene, strikingly high four-phonon scattering rate is found, which is even comparable to the three-phonon scattering rate \cite{feng2018four}. The inclusion of four-phonon scattering in graphene reduces the relative contribution of the out-of-plane acoustic phonon mode (ZA) in $\kappa_{\text{L}}$ from 70\% to 30\%. In case of ML-MoS$_{2}$, the better agreement of the $\kappa_{\text{L}}$ from MD simulations with experiment and the fact that, MD calculations consider all possible phonon scattering events, suggests the possible importance of higher-order anharmonic terms. Therefore, first-principles calculation incorporating the higher-order phonon scattering processes is necessary to understand the thermal transport properties of ML-MoS$_{2}$ accurately.   
	
In earlier studies, certain geometrical and phonon characteristics are identified as the governing conditions behind the significantly high four-phonon scattering strength. For example, acoustic-optical (a-o) frequency gap in BAs \cite{feng2017four}, strong anharmonicity in Lennard-Jones argon \cite{feng2016quantum} and mirror reflection symmetry in graphene \cite{feng2018four}. The presence of reflection symmetry in 2D materials imposes a unique selection rule (RSSR) on the phonon scattering events, i.e., for all orders of phonon-phonon scattering, processes that involve an odd number of out-of-plane phonons (flexural) are forbidden \cite{lindsay2010flexural}. Therefore, three-phonon processes may involve zero or two flexural phonons, while the four-phonon processes may include zero, two or four flexural phonons. In graphene, the RSSR is found to play a decisive role in significantly reducing the relative contribution of the ZA mode in $\kappa_{\text{L}}$ \cite{feng2018four}. Apart from the identified conditions, the quadratic dispersion of the ZA mode, and the resulting large low-frequency phonon population in 2D materials, is expected to play a significant role behind the high four-phonon scattering strength in single-layer graphene and TaS$_{2}$ \cite{feng2018four, zhang2022four}. To understand the coupling between the quadratic phonon dispersion and 4-phonon scattering strength, the associated quadraticity of the ZA phonon mode needs to be altered. However, the modification in the dispersion of the ZA mode has to be done in such a way that the essential conditions, as noted earlier, are still met. For ML-MoS$_{2}$ as well as for ML-ZnO, it is previously reported that with the application of in-plane tensile strain the dispersion of the ZA mode changes from quadratic to linear \cite{chaudhuri2023strain1, chaudhuri2023ab}. The modulation in the 3- and 4-phonon scattering rates with strain, and thereby, with the quadraticity of the ZA branch will provide a clear understanding of the intrinsic phonon scattering mechanism in ML-MoS$_{2}$. It is, therefore, interesting to study the three- and four-phonon scattering processes in ML-MoS$_{2}$ as a function of mechanical strain, which is lacking in literature. One such study may reveal the underlying details of how the various geometrical and phonon dispersion features influence the four-phonon scattering processes in any general material as well.

The thermal transport properties of ML-MoS$_{2}$ has been avidly explored both theoretically and experimentally in the past. However, a wide range of values of the thermal conductivity ($\kappa_{\text{L}}$) prevail in literature. Apart from making the understanding of the thermal transport mechanism ambiguous, this leads to delusive conclusions regarding the performance of ML-MoS$_{2}$ in various applications, such as thermoelectrics. In the present work, we have exemplified the importance of 4-phonon scattering in the thermal transport mechanism of ML-MoS$_{2}$. The accuracy of our calculations are demonstrated by the agreement of the obtained thermal conductivity values with experiments and MD simulations. Besides, we have illustrated the interplay among the conditions that govern the 4-phonon scattering, with particular emphasis on the quadratic dispersion of the ZA phonon mode. Our work, therefore, not only clears the ambiguity regarding the magnitude of lattice thermal conductivity of ML-MoS$_{2}$, but also identify the strong coupling between the phonon dispersion and the 4-phonon scattering processes. Thereby, significantly improves the predicting ability of the first-principles based thermal transport calculations as well as the understanding of the heat transport mechanism of ML-MoS$_{2}$ and any 2D material in general.

\section{Computational details}
First-principles calculations have been performed using ab-initio density functional theory (DFT) as implemented in the Vienna Ab Initio Simulation Package (VASP) \cite{kresse1996efficient, kresse1996efficiency} together with projector augmented wave (PAW) potentials to account for the electron-ion interactions \cite{kresse1999ultrasoft}. The electronic exchange and correlation (XC) interactions are addressed within the generalized gradient approximation (GGA) of Perdew-Burke-Ernzerhof (PBE) \cite{perdew1996generalized}. In all calculations, the Brillouin zone (BZ) is sampled using a well-converged Monkhorst-Pack \cite{monkhorst1976special} k-point set ($ 21\times 21\times 1$), and a conjugate gradient scheme is employed to optimize the geometries until the forces on each atom are less than 0.01 eV/$\text{\AA}$. In the present calculations, in-plane biaxial strain has been applied by isotropically changing the in-plane lattice parameters to a desired value and then, the internal forces acting on each atom is minimized. A vacuum thickness of approximately 20 $\text{\AA}$ is used to avoid the spurious interaction between the periodic images of the layers. 

The phonon dispersion curves are calculated based on the supercell approach using the finite displacement method with an amplitude of 0.015 $\text{\AA}$ as implemented in the phonopy code \cite{togo2015first}. To compute the lattice transport properties, the Boltzmann transport equation (BTE) for phonons is solved as implemented in the ShengBTE code \cite{li2014shengbte}. The second-, third- and fourth-order interatomic force constants (IFC) are calculated based on the finite-difference supercell method. For the third-order IFC a $ 6 \times 6 \times 1$ supercell is created and interactions up to the 6th nearest-neighbour (NN) atom are considered using an extension module of ShengBTE \cite{li2014shengbte}. Similarly, for the fourth-order IFC a $ 4 \times 4 \times 1$ supercell is adopted and 3 NN atoms are considered. Well-converged k-meshes, together with a strict energy convergence criterion of $10^{-8}$ eV, are used in all the supercell based calculations. The three- and four-phonon scattering rates have been computed using ShengBTE \cite{li2014shengbte}. To accurately compute the lattice thermal conductivity, a dense q-mesh of $ 51 \times 51 \times 1$ is used to sample the reciprocal space of the primitive cells. 

Combining the first-principle calculations with the semi-classical Boltzmann transport equation (BTE), the lattice thermal conductivity ($\kappa_{\text{L}}$) of a material can be calculated as:\\
	
\begin{center}

$ \kappa_{\text{L}}^{\text{xy}} = \dfrac{1}{\text{N}_{\text{q}}V} \sum {\hbar}\omega_\lambda {\dfrac{\partial n_{\lambda}^{0}}{\partial T}} \nu_{\lambda , x} \nu_{\lambda , y} \tau_{\lambda} = \dfrac{1}{\text{N}_{\text{q}}V} \sum c_{\lambda} \nu_{\lambda , x} \nu_{\lambda , y} \tau_{\lambda}$ \\

\end{center}
		
where $\lambda$ is a particular phonon mode with wave vector q and frequency $\omega_\lambda$, V is the volume of the Brillouin zone (BZ), N$_{\text{q}}$ corresponds to the total number of q-points sampled in the first BZ, n$_{\lambda}^{0}$ is the Bose-Einstein distribution function associated with the phonon mode $\lambda$ at a particular temperature T, $\nu_{\lambda , x}$ is the x component of the phonon group velocity, $\tau_{\lambda}$ is the carrier lifetime of the $\lambda$ mode phonons. The term $\hbar \omega_\lambda \dfrac{\partial n_{\lambda}^{0}}{\partial T}$ corresponds to the phonon specific heat of the $\lambda$ mode. The intrinsic phonon scattering rates ($\tau_{\lambda}^{-1}$) and thereby, the carrier lifetime ($\tau_{\lambda}$) are computed according to the Fermi’s golden rule and the detailed derivation of the formulas can be found elsewhere \cite{feng2016quantum, feng2017four, feng2018four, yang2019stronger, zhang2022four}. In this work, both the 3- and 4-phonon processes are considered in the scattering rate calculations. Thus, the total scattering rate corresponding to the phonon mode $\lambda$ is computed by the Matthiessen’s rule \cite{mahan2000many}, given as: \\

\begin{center}
		
${\dfrac{1}{\tau_{\lambda}}} = \dfrac{1}{\tau_{3, \lambda}} + \dfrac{1}{\tau_{4, \lambda}}$ \\

\end{center}
 
where $\frac{1}{\tau_{3, \lambda}}$ and $\frac{1}{\tau_{4, \lambda}}$ denote the 3-phonon and 4-phonon scattering rates, respectively. The total scattering rate is, therefore, computed by summing the contribution from all the possible phonon modes.

The phonon BTE in the steady-state, describing the balance of the phonon population for a particular mode $\lambda$ between the phonon diffusion and scattering, is given as,\\

\begin{center}

 $\nu_{\lambda}\cdot \Delta n_{\lambda} = \dfrac{\partial n_{\lambda}}{\partial t}$\\
 
\end{center}
 
where $\nu_{\lambda}$ is the phonon group velocity and $n_{\lambda}$ is the Bose-Einstein occupation distribution of the phonon mode $\lambda$. Within the framework of relaxation time approximation (RTA), including the 3-phonon and 4-phonon scattering processes, the scattering term on the right-hand side can be expressed as: $\dfrac{\partial n_{\lambda}}{\partial t} = \dfrac{n_{\lambda}- n_{\lambda}^{0}}{\tau_{\lambda}}$, where $n_{\lambda}^{0}$ is the equilibrium phonon distribution. Due to the fact that Normal processes (N) contribute indirectly to the thermal resistance by influencing the phonon distribution and thereby, the Umklapp process (U), a direct summation of the N and U scattering rates within the RTA leads to an inaccurate description of the overall thermal transport. Therefore, when N processes are significant as compared to the U processes, an exact solution of the phonon BTE beyond the RTA is required. Within the iterative scheme (ITR) the phonon BTE is solved by considering all the phonon scattering events simultaneously, as opposed to the RTA where scattering events are treated independent to each other. In the ITR, each phonon's final state is coupled with the final states of all other phonons and thus, the current scattering events can be influenced by the results of all the former scattering events. Therefore, the ITR can separate the thermal resistance contribution from each phonon mode.

\section{Results and Discussion}
\subsection{Pristine Monolayer MoS$_{2}$}

\begin{figure}[h!]
	\centering
	\includegraphics[scale=0.3]{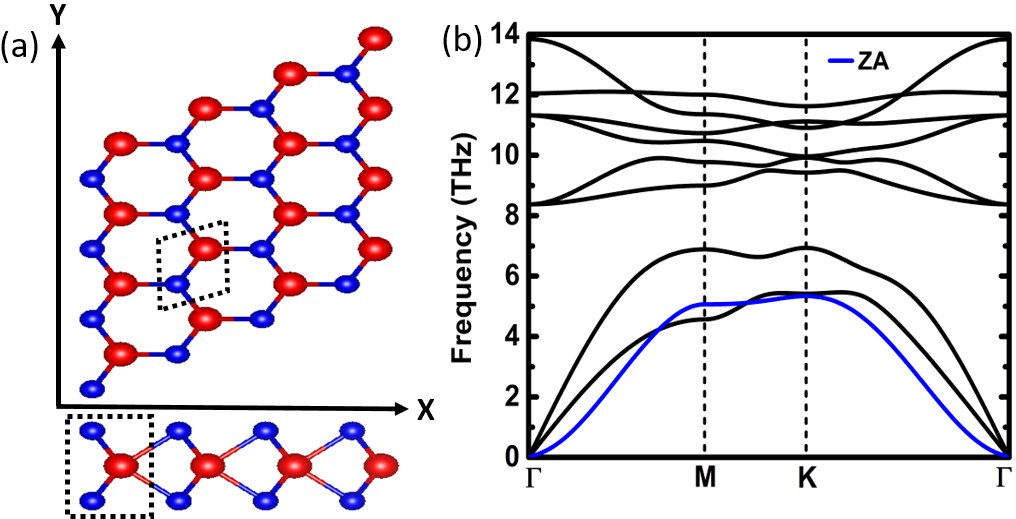}
	\caption{ (a) Crystal structure (top and side view) of monolayer MoS$_{2}$, where the Mo and S atoms are presented as red and blue balls. The black dashed box represents the primitive unit cell. (b) Phonon dispersion of ML-MoS$_{2}$ along the high-symmetry path ($\Gamma$-M-K-$\Gamma$) of the Brillouin zone (BZ). The ZA phonon branch is highlighted in blue. }
	\label{Fig.1}
\end{figure}
	
The single-layer counterpart of 2H-MoS$_{2}$ consists of a Mo atomic plane sandwiched between two S atomic planes. The unit cell of monolayer MoS$_{2}$ in the undistorted 2H phase is composed of a triangular lattice of Mo atoms arranged in a trigonal prismatic structure with the two triangular lattices of S atoms, as can be seen in Fig. \ref{Fig.1}. Structural relaxation results in a lattice parameter of a = b = 3.18 $\text{\AA}$, consistent with the previous reports \cite{chaudhuri2023strain1, jena2017compressive}. Monolayer (ML) 2H-MoS$_{2}$ is a direct band gap semiconductor with a band gap of 1.7 eV \cite{chaudhuri2023strain, chaudhuri2023strain1, jena2017compressive}. Therefore, the heat conduction through ML-MoS$_{2}$ is largely governed by the phonon transport characteristics. From the phonon dispersion of pristine ML-MoS$_{2}$, as shown in Fig. \ref{Fig.1}, certain important characteristics related to the phonon transport and intrinsic phonon-phonon scattering can be immediately identified. For example, the quadratic dispersion of the out-of-plane acoustic mode (ZA) or the flexural mode, which is typical of a 2D material, and the acoustic-optical (a-o) phonon frequency gap of 1.21 THz due to the mass ratio of Mo and S atoms. These characteristics of the phonon dispersion are found to have significant implications over the intrinsic phonon scattering processes in some earlier studies \cite{feng2016quantum, feng2017four, feng2018four, yang2019stronger, zhang2022four}.

\begin{figure}[h!]
	\centering
	\includegraphics[scale=0.45]{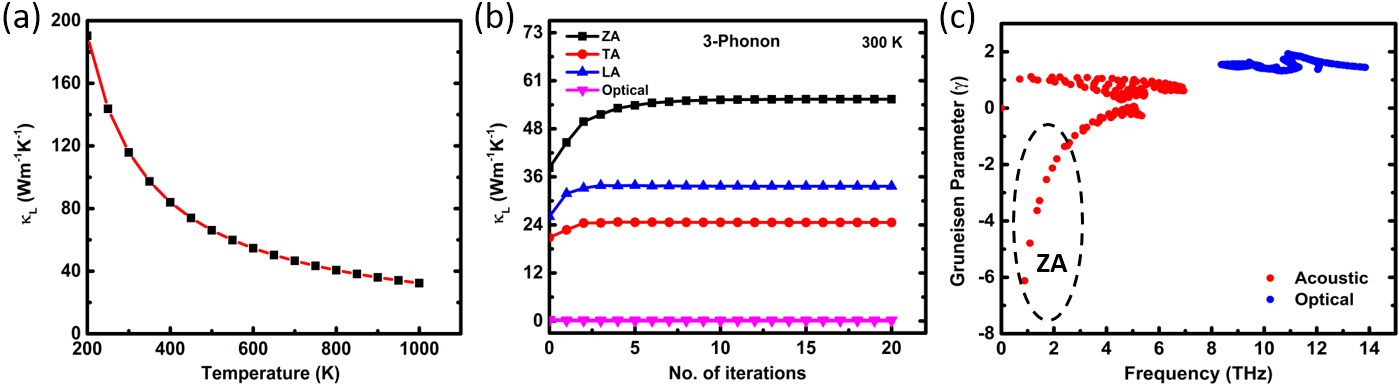}
	\caption{(a) Variation in lattice thermal conductivity ($\kappa_{\text{L}}$) of ML-MoS$_{2}$ with temperature considering only 3-phonon scattering processes. (b) Contribution to the total $\kappa_{\text{L}}$ from the individual phonon modes i.e. the flexural acoustic (ZA), transverse acoustic (TA), longitudinal acoustic (LA) and optical modes at a temperature of 300 K plotted as a function of number of iterations. (c) Variation in the Gr$\ddot{\text{u}}$neisen parameter ($\gamma$) as a function of phonon frequency. The contribution stemming from the ZA mode is highlighted by the oval circle.}
	\label{Fig.2}
\end{figure}
	
The intrinsic phonon scattering directly impacts the lattice thermal transport properties and therefore, the lattice thermal conductivity ($\kappa_{\text{L}}$) can be taken as a measure of the intensity of phonon scattering. Large phonon scattering rate results in a small carrier lifetime ($\tau$) and thereby, leads to a small $\kappa_{\text{L}}$. Our calculated $\kappa_{\text{L}}$ of ML-MoS$_{2}$ considering only 3-phonon scattering within the relaxation time approximation (RTA) as a function of temperature (T) is presented in Fig. \ref{Fig.2}. The value of $\kappa_{\text{L}}$ obtained at 300 K using the relaxation time approximation (RTA) is 115.8 Wm$^{-1}$K$^{-1}$, which agrees well with earlier first-principles calculation based reports \cite{sharma2019electron, gu2014phonon, gandi2016thermal}. A comparison of the $\kappa_{\text{L}}$ values of ML-MoS$_{2}$ at 300 K obtained from experiments and different theoretical works are provided in Table \ref{tab:Table 1}. A large discrepancy in the values of $\kappa_{\text{L}}$ between experiments or MD simulations and first-principle calculations can be seen from Table \ref{tab:Table 1} and also discussed in the Introduction section. This limited accuracy poses serious question towards the approach used in first-principle calculations. For more insights, the contribution towards the $\kappa_{\text{L}}$ at 300 K and the Gr$\ddot{\text{u}}$neisen parameter ($\gamma$) for individual phonon modes are calculated (see Fig. \ref{Fig.2}). It can be seen that the ZA mode contributes the highest to the total $\kappa_{\text{L}}$ and therefore, would have a significantly large relaxation time ($\tau$). In contrast, the magnitude of $\gamma$ associated with the ZA mode is also very high (see Fig. \ref{Fig.2}), which generally corresponds to strong anharmonicity and therefore, high phonon scattering rate ($\tau^{-1}$). In spite of having the highest $\gamma$, such a low scattering rate corresponding to the ZA phonons is something unusual. Clearly, the large overestimation of the total thermal conductivity and the $\tau$ of the ZA phonons suggests the inadequacy of the 3-phonon based first-principle BTE calculations in predicting the thermal transport properties of ML-MoS$_{2}$.

\begin{table}[h]
	\caption{\label{tab:Table 1} Room temperature (300 K) lattice thermal conductivity ($\kappa_{\text{L}}$) of ML-MoS$_2$ obtained in this work and from literatures using different levels of approximations.}
	\begin{tabular*}{0.55\textwidth}{| c | @{\extracolsep{\fill}} c | c |}
		\hline
		Reference & Method & $\kappa_{\text{L}}$ (Wm$^{-1}$K$^{-1}$) \\
		\hline
		This work & DFT: 3-phonon (RTA) & 115.8 \\
		\hline
		Ref. \cite{yan2014thermal} & Experiment: monolayer & 34.5 \\
		\hline
		Ref. \cite{sahoo2013temperature} & Experiment: few-layer & 52 \\
		\hline
		Ref. \cite{sharma2019electron} & DFT: 3-phonon (RTA) & 116.8 \\  
		\hline 
		Ref. \cite{gandi2016thermal} & DFT: 3-phonon (RTA) & 131 \\
		\hline
		Ref. \cite{krishnamoorthy2019thermal} & MD & 38.1 \\
		\hline
		Ref. \cite{wei2014phonon} & MD & 26.2 \\  
		\hline
		Ref. \cite{cai2014lattice} & DFPT + NEGF & 23.2 \\ 
		\hline
		This work & DFT: 3-phonon + 4-phonon (RTA) & 24.6 \\
		\hline
		This work & DFT: 3-phonon + 4-phonon (ITR) & 27.7 \\
		\hline
	\end{tabular*}
\end{table}

\begin{figure}[h!]
	\centering
	\includegraphics[scale=0.3]{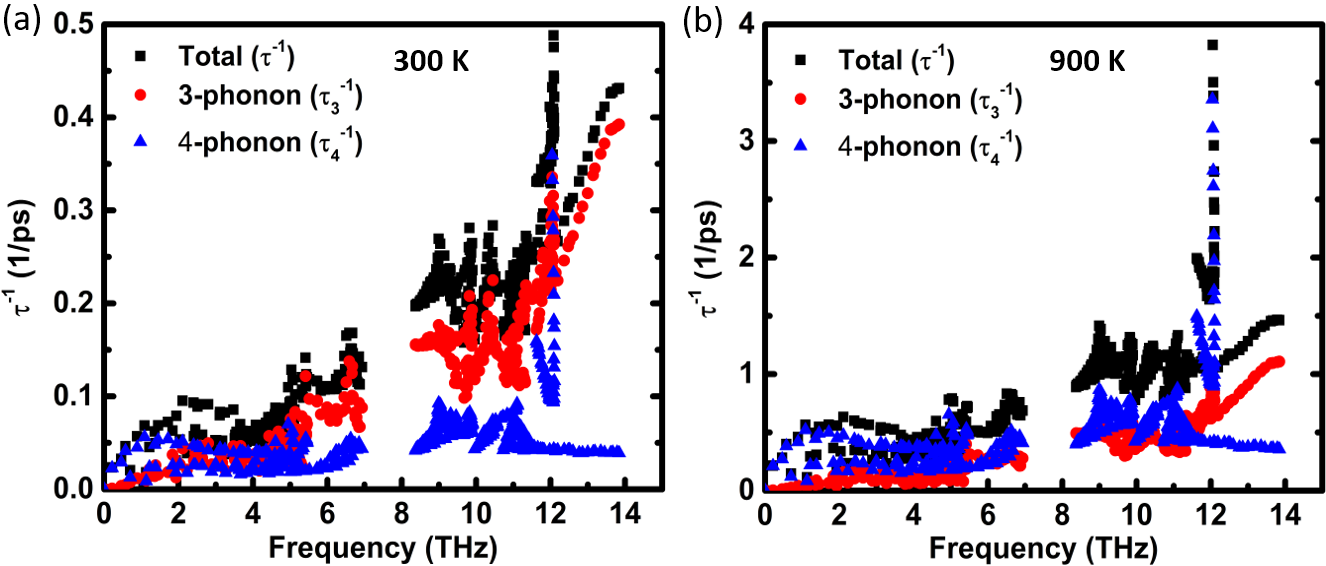}
	\caption{ Variation in the 3-phonon ($\tau_{3}^{-1}$; red dots), 4-phonon ($\tau_{4}^{-1}$; blue dots) and total ($\tau_{3}^{-1}$ + $\tau_{4}^{-1}$; black dots) scattering rates of ML-MoS$_{2}$ with phonon frequency at (a) 300 K and (b) 900 K. }
	\label{Fig.3}
\end{figure}

To understand the role of higher order anharmonicity in the phonon scattering and therefore, the transport properties, we calculate the 4-phonon scattering rate of ML-MoS$_{2}$. The variation in the 3- and 4-phonon scattering rates of ML-MoS$_{2}$ as a function of phonon frequency is calculated at two different temperatures, 300 K and 900 K, and presented in Fig. \ref{Fig.3}. It can be seen that, at 300 K the 4-phonon scattering rate is comparable and even higher compared to the 3-phonon scattering rates at some frequencies, especially for the acoustic phonons. At higher temperatures, such as 900 K, the 4-phonon scattering dominates the overall phonon scattering processes. Such a high 4-phonon scattering rate is against the general notion of perturbation theory, which states that the strength of higher-order scattering depends on the magnitude of corresponding higher-order terms of the Hamiltonian, which gets progressively smaller with the order number. Ruan $\textit {et al.}$ have demonstrated a similar phenomenon in a number of materials \cite{feng2016quantum, feng2017four, feng2018four, yang2019stronger}. Notably, the low frequency regime, such as 0-2 THz, the 4-phonon scattering rate is astonishingly high and the 3-phonon scattering rate is negligible. This can be correlated with the unique quadratic dispersion of the flexural (ZA) phonons near the zone center ($\Gamma$). In the frequency range of 0-2 THz, the ZA phonons have the highest population owing to the quadratic dispersion and therefore, the corresponding scattering phase space is also high.

The low 3-phonon scattering rate of the ZA phonons can be understood due to the simultaneous effect of a-o frequency gap and the selection rule imposed by the reflection symmetry (RSSR). Owing to the horizontal mirror reflection symmetry of ML-MoS$_{2}$, of which the mirror plane is in the Mo atomic plane, the RSSR is relevant to all orders of phonon scattering processes. This unique phenomenon in 2D materials is found to forbid nearly 60-90\% 3-phonon scattering processes of the ZA mode in graphene and thereby, results in a large overestimation of the $\kappa_{\text{L}}$ associated with it \cite{feng2018four}. Even in a quasi-two-dimensional material like diamane the horizontal reflection symmetry is found to strongly influence the intrinsic phonon scattering rates \cite{zhu2019giant}. Due to the large a-o frequency gap a large number of 3-phonon processes involving the low-lying ZA phonons and high energy optical phonons are restricted owing to the energy conservation criteria (ECC). This coupled with the RSSR, which prevents scattering events involving odd number of ZA phonons, such as aoo, aaa etc., results in an ultra-high lifetime ($\tau$) corresponding to the ZA mode. However, in the 4-phonon scattering, with the aid of the one extra phonon the a-o gap can be overcame and therefore, the coupling between the acoustic and optical phonons will strengthen. Also, the RSSR allows zero, two or four number of out-of-plane phonons to take part in the scattering events in a 4-phonon process, while only combinations of zero and two phonons are allowed in 3-phonon process. Therefore, the scattering phase space of the ZA phonons in 3-phonon processes is much smaller compared to that of the 4-phonon processes, which results in the superiority of the 4-phonon scattering rates in the low frequency range.

\begin{figure}[h!]
	\centering
	\includegraphics[scale=0.35]{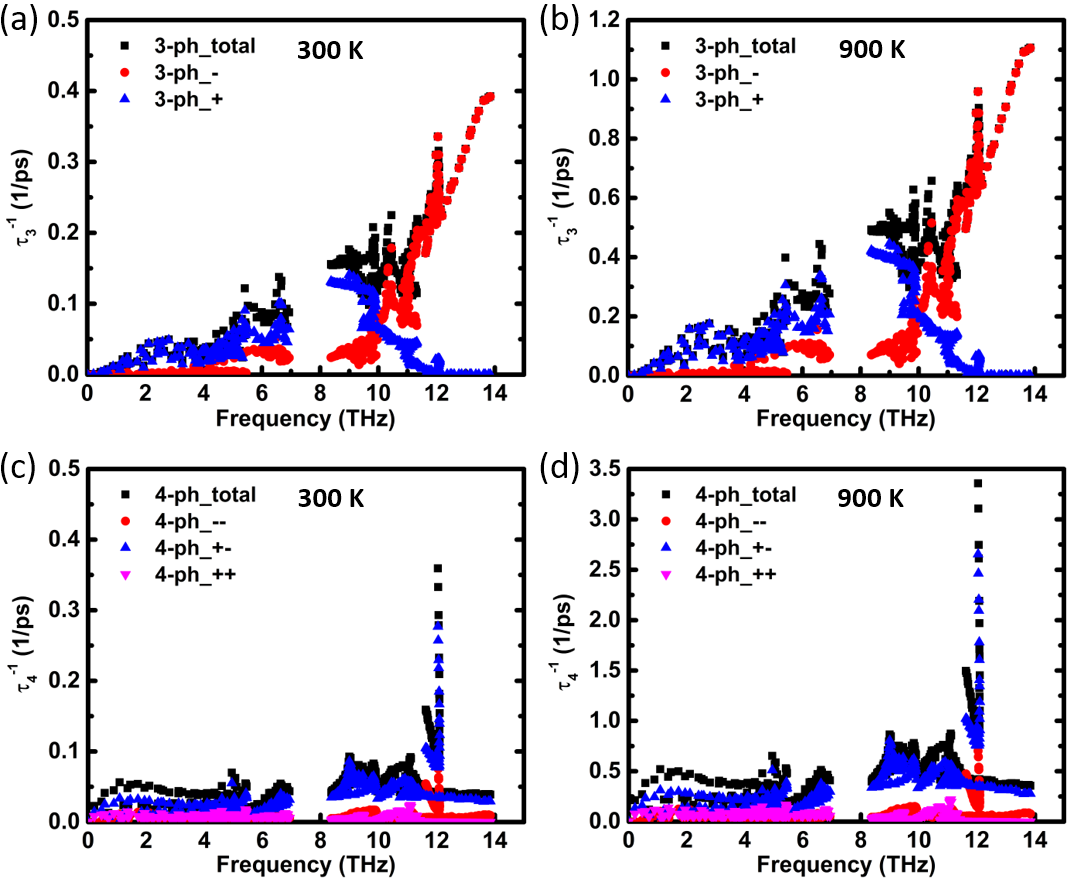}
	\caption{ Variation in the scattering rates corresponding to the (a, b) 3-phonon and (c, d) 4-phonon scattering processes at 300 K and 900 K. The 3-phonon scattering processes are denoted as; ``+": combination, where two phonons combine to form one new phonon and ``-": splitting, where one phonon splits into two phonons. The 4-phonon scattering events are denoted as; ``++": combination, where three phonons combine to form a single phonon, ``--": splitting, where one phonon splits into three phonons and ``+-": redistribution, where two phonons scatter to generate two new phonons.}
	\label{Fig.4}
\end{figure}

To get further insight of the scattering processes, we calculate the scattering rates corresponding to every possible ways of scattering allowed in the 3- and 4-phonon processes. There are two possibilities in a scattering process involving three phonons, such as, a single phonon may split into two phonons (splitting, -), or two phonons may combine to create a new phonon (combination, +). Similarly, four phonons may engage in three possible ways, such as, a single phonon may split into three phonons (splitting, --), or three phonons may combine to form a new phonon (combination, ++), or two phonons may combine to create two new phonons (redistribution, +-). The variation in the scattering rates corresponding to all the possible channels of 3- and 4-phonon processes at temperatures 300 K and 900 K are presented in Fig. \ref{Fig.4}. For the 3-phonon processes, in the low energy region (0-7 THz) the three phonon combination (+) is favorable, while in the high energy region (8-14 THz) the splitting (-) is dominant. This can be understood as a consequence of the energy conservation criteria (ECC). For the 4-phonon processes, on the other hand, the four phonon redistribution (+-) is found to be the dominant one in the entire frequency range. Due to the nearly dispersionless character of the ZO$_{2}$ mode at a frequency around 12 THz, the 4-phonon scattering rate, albeit mainly the four phonon redistribution process peaks.

\begin{figure}[h!]
	\centering
	\includegraphics[scale=0.4]{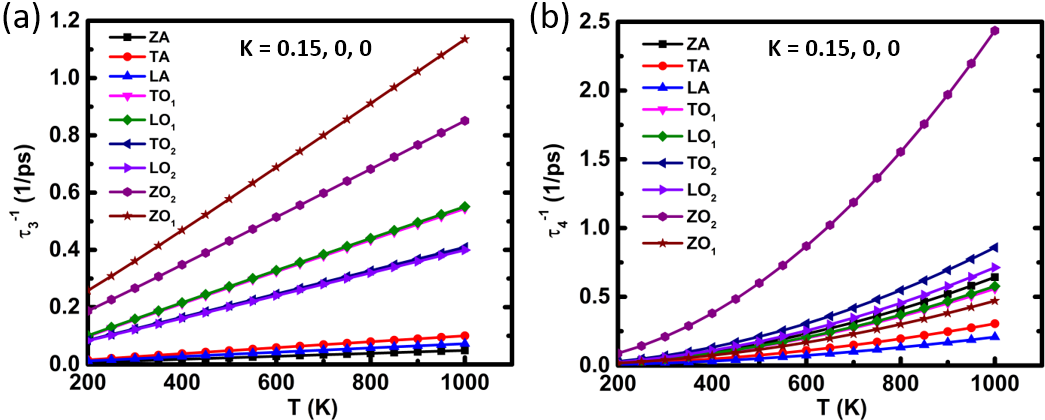}
	\caption{ (a) 3-phonon and (b) 4-phonon scattering rates plotted as a function of temperature corresponding to all the phonon modes of ML-MoS$_{2}$. The scattering rates are calculated at a particular k-point (0.15, 0, 0) of the BZ located between $\Gamma$ and M points.}
	\label{Fig.5}
\end{figure}

From Fig. \ref{Fig.3} and \ref{Fig.4} it is clear that the four phonon scattering, although significant at room temperature, become much more important at higher temperatures. To understand the superiority of the 4-phonon scattering at high temperatures, we calculate the temperature dependency of the phonon scattering rates corresponding to all the nine phonon modes at a particular k-point (0.15, 0, 0) of the BZ, and shown in Fig. \ref{Fig.5}. It can be seen that while $\tau_{3}^{-1}$ increases linearly with temperature, $\tau_{4}^{-1}$ increases quadratically. These temperature dependencies follow from the fact that, at a particular temperature $\tau_{3}^{-1}$ is proportional to the phonon population, whereas the $\tau_{4}^{-1}$ depends on the square of the population, and the phonon population scales linearly with temperature. Such scaling laws of the scattering rates have been seen in earlier reports \cite{feng2016quantum, feng2017four} as well, suggesting the validity of our phonon scattering calculations. From Fig. \ref{Fig.5}, it is clear that the $\tau_{3}^{-1}$ and $\tau_{4}^{-1}$ associated with the optical modes are the highest and therefore, their contribution towards the thermal transport is significantly less compared to the acoustic modes. Amongst the three acoustic modes, the ZA mode is found to have the lowest $\tau_{3}^{-1}$ in the entire temperature range (see Fig. \ref{Fig.5}(a)). However, the $\tau_{4}^{-1}$ corresponding to the ZA mode is remarkably high compared to the other two acoustic modes and even higher compared to some of the optical modes. The highest values of $\tau_{4}^{-1}$, at all temperatures, correspond to the ZO$_{2}$ mode due to the nearly flat bandstructure and high availability of phononic states, as has been discussed earlier. To further support the above conclusions, the calculation has been repeated at another k-point (0.3, 0, 0) and similar phenomena is observed, as can be seen in Fig. S1 (see Supplementary Information).

\begin{figure}[h!]
	\centering
	\includegraphics[scale=0.4]{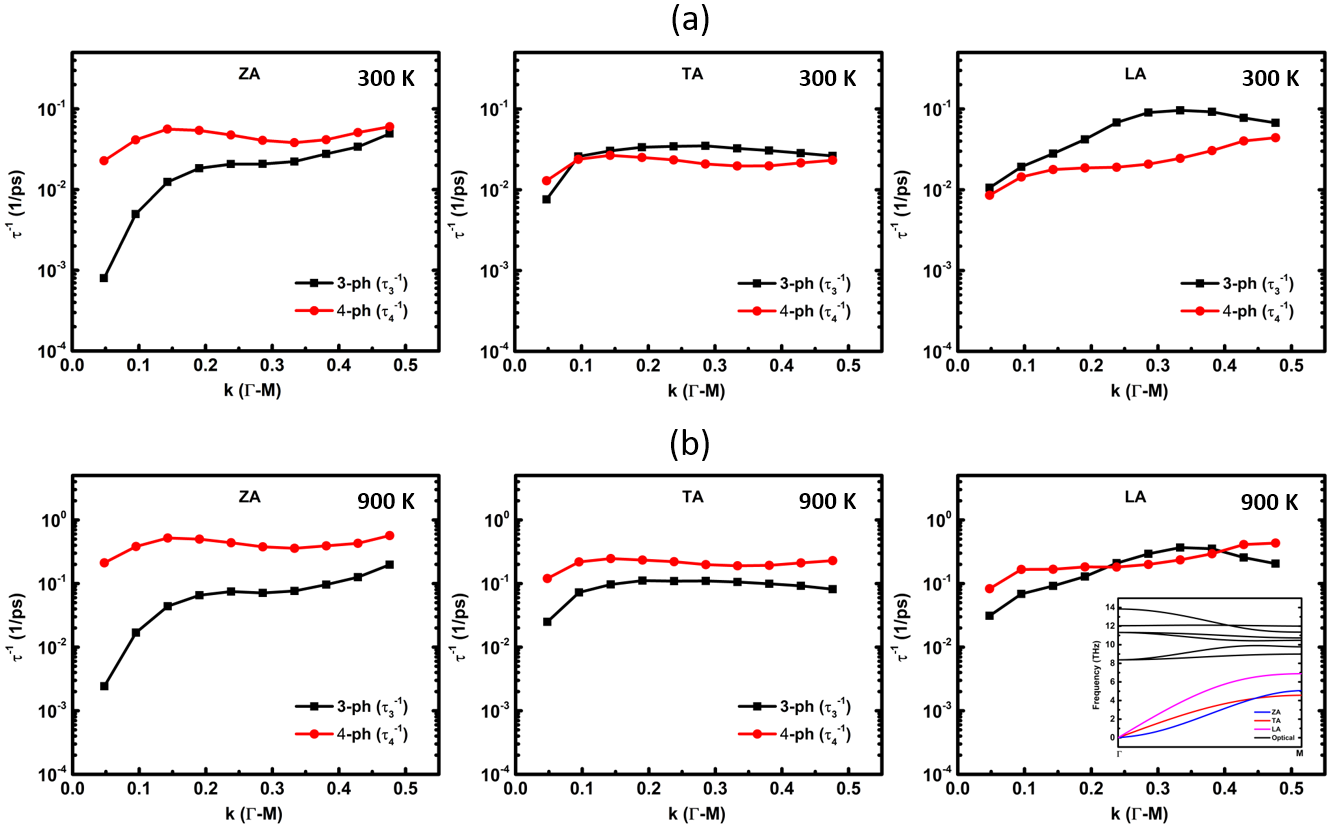}
	\caption{ Variation in 3-phonon ($\tau_{3}^{-1}$) and 4-phonon ($\tau_{4}^{-1}$) scattering rates corresponding to the three acoustic modes (ZA, TA and LA) plotted as a function of the reduced wave vector along the high symmetry k-path of $\Gamma$-M at a temperature of (a) 300 K and (b) 900 K. The x-coordinate of 0.0 corresponds to the $\Gamma$-point and 0.5 corresponds to the M point of the BZ. The phonon dispersion of ML-MoS$_{2}$ from $\Gamma$ to M is shown in inset.}
	\label{Fig.6}
\end{figure}

To understand the phonon scattering mechanism of the acoustic modes in detail, we calculate the $\tau_{3}^{-1}$ and $\tau_{4}^{-1}$ with respect to the reduced wave vector along a specific k-path ($\Gamma$-M) of the BZ for the three acoustic modes (ZA, LA and TA) at temperatures 300 K and 900 K, and presented in Fig. \ref{Fig.6}. We chose to calculate the scattering rates only for the acoustic modes as the thermal transport properties of ML-MoS$_{2}$ are almost entirely governed by the acoustic modes. It can be seen that at T = 300 K, except for the ZA modes, $\tau_{3}^{-1}$ is higher than the $\tau_{4}^{-1}$ throughout the k-path, which is aligned with the general notion following the perturbation theory. However, for the ZA mode, the $\tau_{4}^{-1}$ is well above $\tau_{3}^{-1}$ for all k-points in the high-symmetry path $\Gamma$-M. The difference, $\tau_{4}^{-1}$ - $\tau_{3}^{-1}$, is large for k-points near the zone center ($\Gamma$-point), which corresponds to the low frequency region. At higher temperatures, such as 900 K, due to the quadratic dependency of $\tau_{4}^{-1}$ on temperature, the 4-phonon process dominates and $\tau_{4}^{-1}$ becomes higher than $\tau_{3}^{-1}$ for all three acoustic modes. The scattering rates corresponding to all possible channels involving three phonons (+, -) and four phonons (++, +-, --) as a function of the wave vector along $\Gamma$-M is provided in Fig. S2 and S3 (see Supplementary Information). It is found that, while the combination (+) process dominates the 3-phonon processes, the 4-phonon processes are dominated by the redistribution (+-) process. These findings corroborate with what have been observed earlier (see Fig. \ref{Fig.4}) and thus, indicate the accuracy of our calculation.

\begin{figure}[h!]
	\centering
	\includegraphics[scale=0.35]{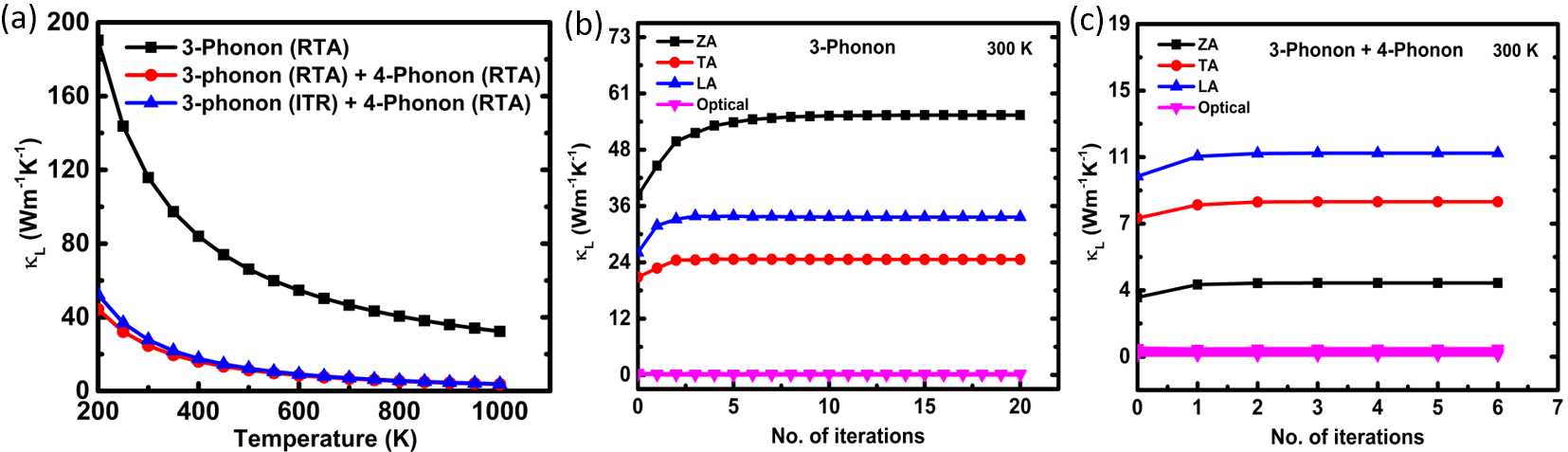}
	\caption{ (a) Variation in the lattice thermal conductivity ($\kappa_{\text{L}}$) of ML-MoS$_{2}$ plotted as a function of temperature. The plot includes results obtained without and with the incorporation of 4-phonon scattering processes with RTA and ITR solution. Contribution in $\kappa_{\text{L}}$ from the individual phonon modes at 300 K using (b) 3-phonon only and (c) both 3- and 4-phonon scattering rates.}
	\label{Fig.7}
\end{figure}

To understand the importance of 4-phonon scattering in the thermal transport properties of ML-MoS$_{2}$, we calculate the lattice thermal conductivity: $\kappa_{\text{L(3)}}$ and $\kappa_{\text{L(3+4)}}$ based on the RTA solved $\tau_{3}^{-1}$ and $\tau_{3}^{-1}$ + $\tau_{4}^{-1}$, respectively. The variation in $\kappa_{\text{L(3)}}$ and $\kappa_{\text{L(3+4)}}$ of ML-MoS$_{2}$ as a function of temperature is shown in Fig. \ref{Fig.7}. We find that the inclusion of 4-phonon scattering into the thermal transport calculations significantly reduces the thermal conductivity from 115.8 Wm$^{-1}$K$^{-1}$ to 24.6 Wm$^{-1}$K$^{-1}$ at temperature 300 K. Our result of $\kappa_{\text{L(3+4)}}$ agrees well with previously reported experimental values and also with the predictions based on MD simulations \cite{yan2014thermal, sahoo2013temperature, krishnamoorthy2019thermal, wei2014phonon, ding2015plane, cai2014lattice}. The $\kappa_{\text{L(3)}}$ computed by accounting only the 3-phonon scattering is seem to significantly overestimate the thermal conductivity as the subsequent incorporation of 4-phonon scattering bring reductions at all temperatures. It is, therefore, clear that, the 4-phonon processes have a compelling role in governing the thermal transport properties of ML-MoS$_{2}$ at all temperatures. To understand the impact of the 4-phonon scattering in greater detail, we calculate the contribution of each phonon mode in $\kappa_{\text{L(3+4)}}$, and presented in Fig. \ref{Fig.7}. The reduction in $\kappa_{\text{L}}$, following the inclusion of 4-phonon scattering, mainly comes from the ZA branch, whose contribution in $\kappa_{\text{L}}$ reduces from $\approx$ 64\% to $\approx$ 16\%. The restricted 3-phonon phase space of the ZA mode, resulting from the a-o frequency gap and RSSR, leads to a remarkably high $\tau$ and thereby, a largely overestimated $\kappa_{\text{L(3)}}$. The large $\tau_{4}^{-1}$ corresponding to the ZA mode suppresses its contribution in $\kappa_{\text{L(3+4)}}$ to a value lower than the other two acoustic modes. The contribution from the optical modes in heat transport remain insignificant even after the incorporation of 4-phonon scattering.

Additionally, we observed that our result of $\kappa_{\text{L(3+4)}}$ (24.6 Wm$^{-1}$K$^{-1}$) using the relaxation time approximation (RTA) is less than the experimentally measured value. This is something abnormal, since in experiments a phonon, apart from the scattering with another phonon, may scatter from various other sources, such as defects, grain boundary etc. depending on the material parameters. The under-prediction in the values of $\kappa_{\text{L(3+4)}}$ is found to originate from the fact that, in RTA calculations both the Normal (N) and Umklapp (U) processes are considered to be a source of direct thermal resistance. However, only the U processes can introduce thermal resistance directly, whereas the N processes are generally considered as non-resistive. Therefore, the RTA can produce accurate prediction of $\kappa_{\text{L}}$ only for those materials where N processes are negligible. It can be seen from Fig. S4 (see Supplementary Information) that, in case of ML-MoS$_{2}$, N processes dominate the overall phonon scattering, similar to graphene [30]. From the variation of the scattering rates corresponding to the N and U processes as a function of the wave vector along $\Gamma$-M (see Fig. S5 in Supplementary Information), it is clear that U processes dominate only around the BZ edge. Therefore, appropriate handling of the N and U processes beyond the RTA is important for the accurate prediction of $\kappa_{\text{L}}$. Thus, we have calculated the $\kappa_{\text{L}}$ beyond the RTA by iteratively solving the BTE (ITR) mixing 3-phonon interactions. However, due to the high computational cost, the 4-phonon scattering rates are calculated within the RTA level only and inserted into the iterative scheme. Similar approach has already been used in earlier reports \cite{feng2016quantum, feng2017four}. The ITR solved $\kappa_{\text{L(3+4)}}$ of ML-MoS$_{2}$ at 300 K is found to be 27.7 Wm$^{-1}$K$^{-1}$, which is even more closer to the experimentally found $\kappa_{\text{L}}$ than the RTA predicted one. We, therefore, conjecture that the best accurate values of $\kappa_{\text{L}}$ of ML-MoS$_{2}$ can be found if the 4-phonon processes are incorporated within the iterative scheme. However, compared to the $\approx$ 230\% error in $\kappa_{\text{L(3)}}$, a prediction of $\kappa_{\text{L(3+4)}}$ which is within $\approx$ 20\% of the experimental value, proves the importance of 4-phonon scattering in ML-MoS$_{2}$. A table containing the values of $\kappa_{\text{L}}$ at 300 K at different level of approximations is presented in Table \ref{tab:Table 1}.

\subsection{Strain Induced Modifications}

\begin{figure}[h!]
	\centering
	\includegraphics[scale=0.35]{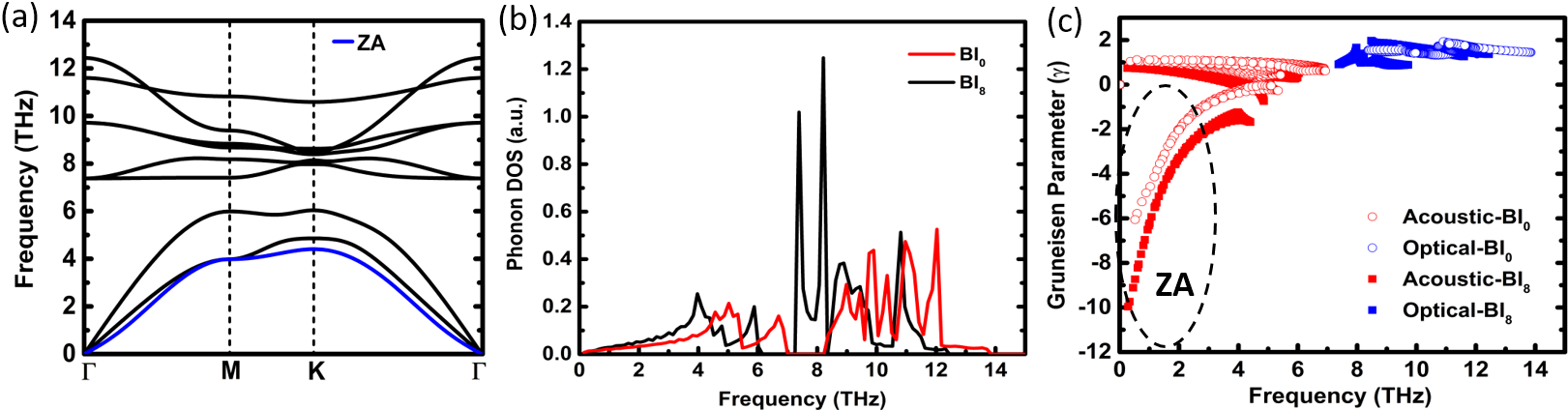}
	\caption{ (a) The phonon dispersion of ML-MoS$_{2}$ under 8\% biaxial tensile strain. The ZA phonon branch is highlighted in blue. Variation in the (b) phonon density of states (DOS) and (c) Gr$\ddot{\text{u}}$neisen parameter ($\gamma$) as a function of phonon frequency in the unstrained (BI$_{0}$) and 8\% biaxially strained (BI$_{8}$) condition.}
	\label{Fig.8}
\end{figure}
	
In order to understand in depth the role of certain geometrical and phonon dispersion characteristics, such as the a-o frequency gap, RSSR, quadratic dispersion of the ZA phonon branch and reflection symmetry etc., in governing the intensity of 4-phonon scattering, we applied in-plane biaxial tensile strain on ML-MoS$_{2}$. The phonon dispersion, density of states (DOS) and the Gr$\ddot{\text{u}}$neisen parameter ($\gamma$) of ML-MoS$_{2}$ under 8\% biaxial tensile strain (BI$_{8}$) along with those in the unstrained condition are shown in Fig. \ref{Fig.8}. Certain changes in the phonon dispersion characteristics induced by the strain can be immediately noticed from Fig. \ref{Fig.8}. For example, hardening of the ZA mode dispersion, which alters the quadratic dispersion in the unstrained condition (BI$_{0}$) into a nearly linear one. The application of strain induces blue shift in the optical phonon modes and softens the LA and TA modes, which results from the weakening of the Mo-S bonding strength. However, the a-o frequency gap remains nearly unchanged in the strained condition: 1.21 THz in the unstrained condition and 1.15 THz in the 8\% biaxial strained case (BI$_{8}$), as can be seen from the phonon DOS in Fig. \ref{Fig.8}. Also, with the application of strain, the bond anharmonicity associated with MoS$_{2}$ increases, as can be seen from the increase in the Gr$\ddot{\text{u}}$neisen parameter ($\gamma$) (see Fig. \ref{Fig.8}(c)). The enhancement in the $\gamma$, especially in the low frequency region (0-2 THz), clearly indicates an enhanced scattering of the acoustic phonons. The application of isotropic in-plane biaxial strain does not alter the hexagonal crystal symmetry of ML-MoS$_{2}$, and therefore, the RSSR, which is an outcome of the mirror reflection symmetry, is still applicable in the strained condition. We understand that the strain induced changes in the phonon specific heat and group velocity can impact the thermal transport properties, and thus, incorporated in the calculation of $\kappa_{\text{L}}$. However, they are not shown here, since the focus of this work is to study the modifications in the intrinsic phonon scattering rates and their impact on the thermal transport properties.

\begin{figure}[h!]
	\centering
	\includegraphics[scale=0.4]{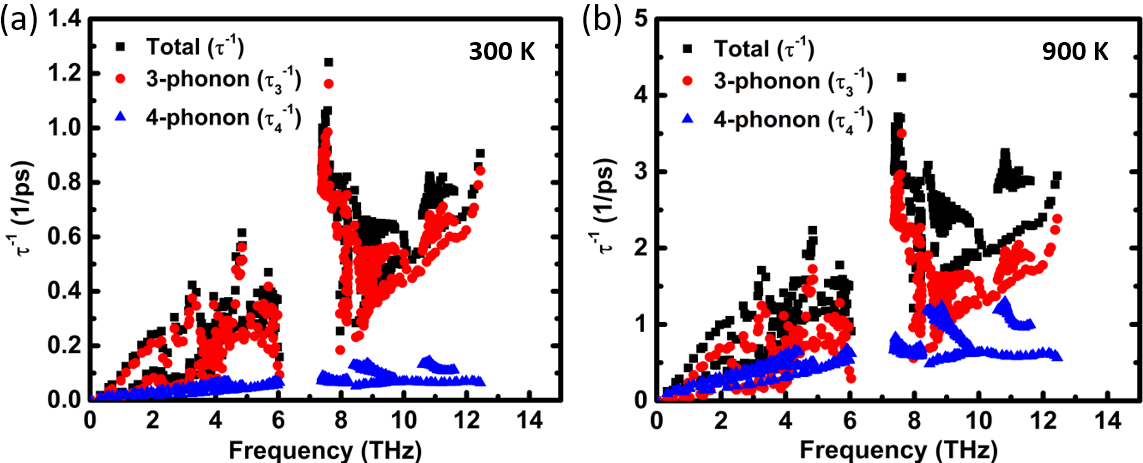}
	\caption{ The 3-phonon ($\tau_{3}^{-1}$; red dots), 4-phonon ($\tau_{4}^{-1}$; blue dots) and total ($\tau_{3}^{-1}$ + $\tau_{4}^{-1}$; black dots) scattering rates of ML-MoS$_{2}$ in the strained condition (BI$_{8}$) at temperatures (a) 300 K and (b) 900 K.}
	\label{Fig.9}
\end{figure}

The features that were responsible for the unusually high 4-phonon scattering in the unstrained condition, such as the a-o frequency gap, RSSR and strong anharmonicity, are all present in the strained condition. It is therefore legitimate to expect a likewise dominance of the 4-phonon scattering over the thermal transport properties of ML-MoS$_{2}$ in the strained condition. However, the $\tau_{4}^{-1}$ in the strained case is found to be negligible compared to the $\tau_{3}^{-1}$ at 300 K, and even at higher temperatures, such as 900 K, remain insignificant. The variation in the phonon scattering rates as a function of frequency at temperatures 300 K and 900 K under 8\% biaxial tensile strain (BI$_{8}$) are presented in Fig. \ref{Fig.9}. It can be seen that, even in the low frequency region (0-2 THz) the scattering strength corresponding to 4-phonon process is inconsequential. In the unstrained condition, on the other hand, the $\tau_{4}^{-1}$/$\tau_{3}^{-1}$ was found to be the highest in the same low frequency region (0-2 THz). The small $\tau_{4}^{-1}$ in the strained condition, having met all the conditions for strong 4-phonon scattering, is surprising and therefore, needs detailed investigation. The scattering rates corresponding to all possible 3-phonon (+, -) and 4-phonon (++, +-, --) processes are computed as a function of the phonon frequency and shown in Fig. S6 (see Supplementary Information). Similar to the unstrained condition, the combination process (+) is found to dominate the 3-phonon processes in the low frequency region, and splitting process (-) in the high frequency region. Similarly, in the 4-phonon process the redistribution process (+-) is observed to dominate in the entire frequency range. It is, therefore, clear that the scattering mechanisms involving three or four phonons are identical in both unstrained and strained cases.

\begin{figure}[h!]
	\centering
	\includegraphics[scale=0.4]{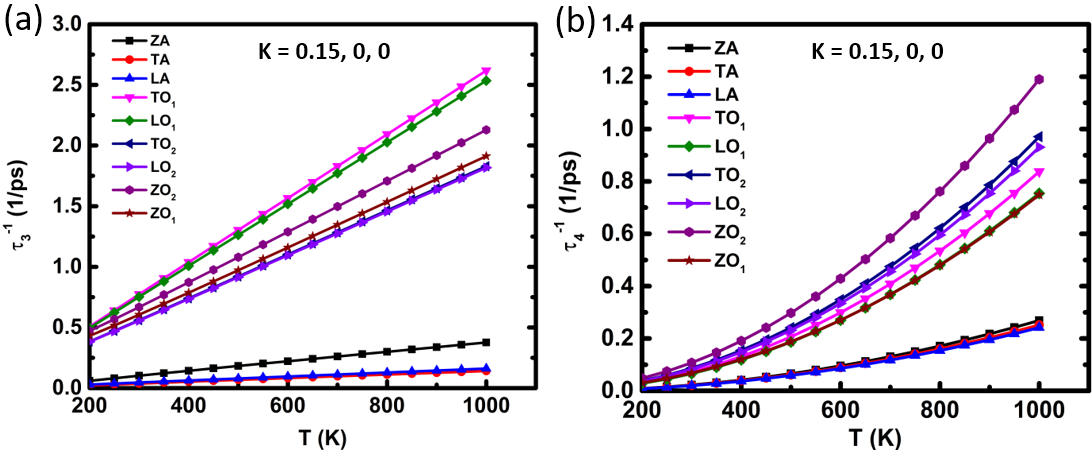}
	\caption{(a) 3-phonon and (b) 4-phonon scattering rates plotted as a function of temperature for different phonon modes in ML-MoS$_{2}$ in the 8 \% biaxially strained condition (BI$_{8}$). The scattering rates are calculated at a particular k-point (0.15, 0, 0) of the BZ located between $\Gamma$ and M points.}
	\label{Fig.10}
\end{figure}

The variation in the $\tau_{3}^{-1}$ and $\tau_{4}^{-1}$ with temperature in a range of 200 K to 1000 K at a particular k-point (k = 0.15, 0, 0) of the BZ are shown in Fig. \ref{Fig.10}. It is clear that the overall scattering rate increases with the application of strain owing to the increase in Gr$\ddot{\text{u}}$neisen parameter ($\gamma$) and the resulting anharmonicity. The magnitude of $\tau_{3}^{-1}$ averaged for all the modes increases $\approx$ 2 times in the strained condition (BI$_{8}$) compared to that of the unstrained one (BI$_{0}$) at nearly all temperatures. However, the mode averaged $\tau_{4}^{-1}$ remain insensitive to the application of strain and thus, observed to be nearly same in both the cases.  The scattering rates corresponding to the optical modes are much higher compared to the acoustic modes for both 3- and 4-phonon processes at all temperatures. Therefore, the role of the optical modes in the lattice thermal transport is insignificant even in the strained case. Amongst the three acoustic modes, the ZA mode, which had the lowest $\tau_{3}^{-1}$ associated with it in the unstrained case, has now the highest $\tau_{3}^{-1}$ in the strained condition. The $\tau_{4}^{-1}$ corresponding to the three acoustic modes are nearly equal in the strained case (BI$_{8}$), and is surprisingly less for the ZA mode as compared to the unstrained case (BI$_{0}$) in the entire temperature range.

\begin{figure}[h!]
	\centering
	\includegraphics[scale=0.4]{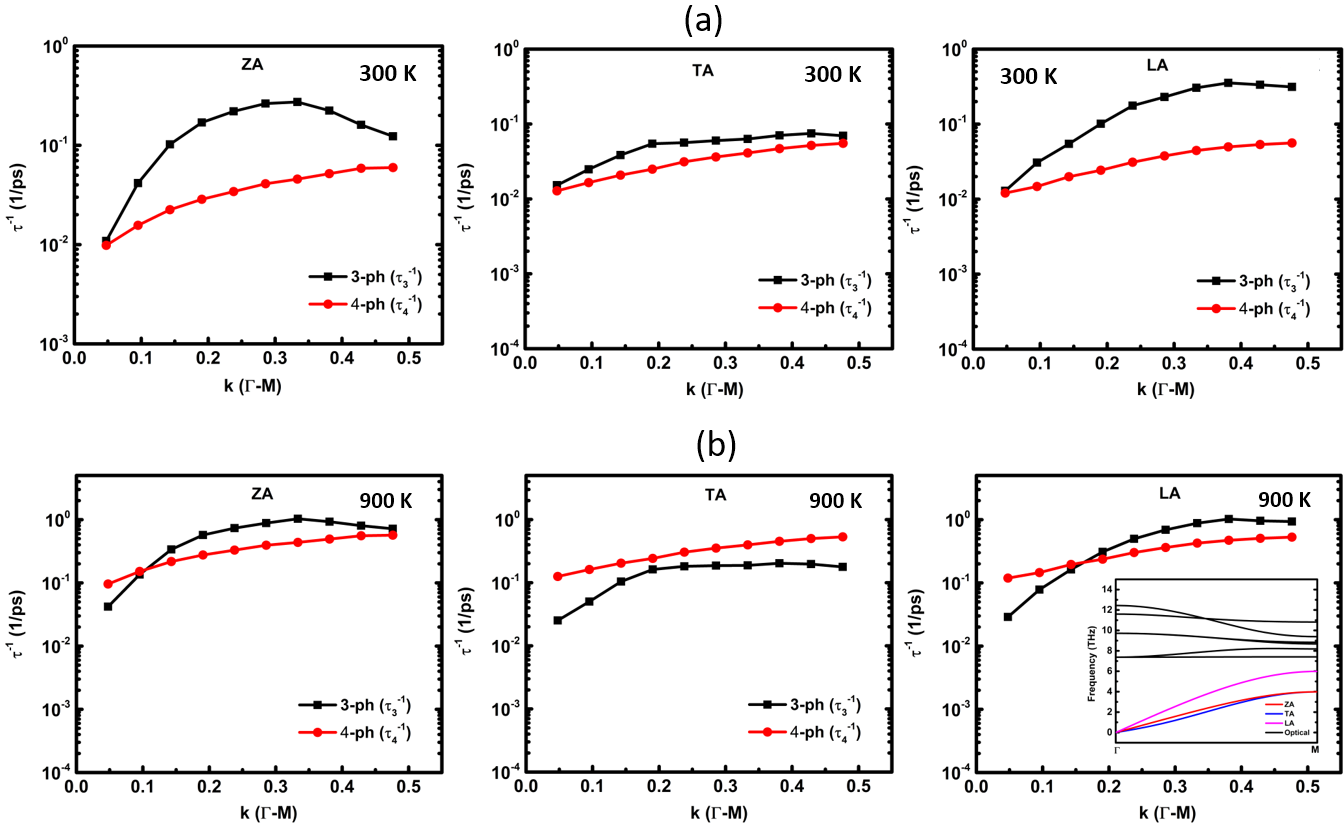}
	\caption{ Variation in 3-phonon ($\tau_{3}^{-1}$) and 4-phonon ($\tau_{4}^{-1}$) scattering rates corresponding to the three acoustic modes (ZA, TA and LA) of ML-MoS$_{2}$ in the strained condition as a function of the reduced wave vector along the k-path $\Gamma$-M at temperatures (a) 300 K and (b) 900 K. The phonon dispersion of 8\% biaxially strained ML-MoS$_{2}$ along $\Gamma$-M is shown in inset.}
	\label{Fig.11}
\end{figure}

To get further insights of the intrinsic phonon scattering processes in the strained ML-MoS$_{2}$, we calculate the $\tau_{3}^{-1}$ and $\tau_{4}^{-1}$ of the acoustic modes along the k-path $\Gamma$-M, as has been done in the unstrained case. The variation in the scattering rates as a function of the wavevector along $\Gamma$-M is presented in Fig. \ref{Fig.11}. Contrary to what is seen in the unstrained condition, the $\tau_{3}^{-1}$ is found to be higher compared to the $\tau_{4}^{-1}$ for all three acoustic modes (ZA, TA and LA) for the entire k-path ($\Gamma$-M) at 300 K. The highest values of the scattering rate ratio ($\tau_{3}^{-1}$/$\tau_{4}^{-1}$) correspond to the ZA mode. In summary, among the three acoustic modes, while the room temperature $\tau_{3}^{-1}$ is lowest and $\tau_{4}^{-1}$ is highest for the ZA mode in the unstrained condition, becomes the exact opposite, i.e. $\tau_{3}^{-1}$ is highest and $\tau_{4}^{-1}$ is lowest, in the strained case. Even at higher temperatures, such as 900 K, the $\tau_{3}^{-1}$ is found to have higher values compared to the $\tau_{4}^{-1}$ for the acoustic modes, except for the TA mode. The variation in the scattering rates of the acoustic modes corresponding to all the possible processes involving three (+, -) or four (++, +-, --) phonons along the k-path $\Gamma$-M are shown in Fig. S7 and S8 (see Supplementary Information). Similar results to what have been observed in Fig. S2 and S3 are seen.

\begin{figure}[h!]
	\centering
	\includegraphics[scale=0.35]{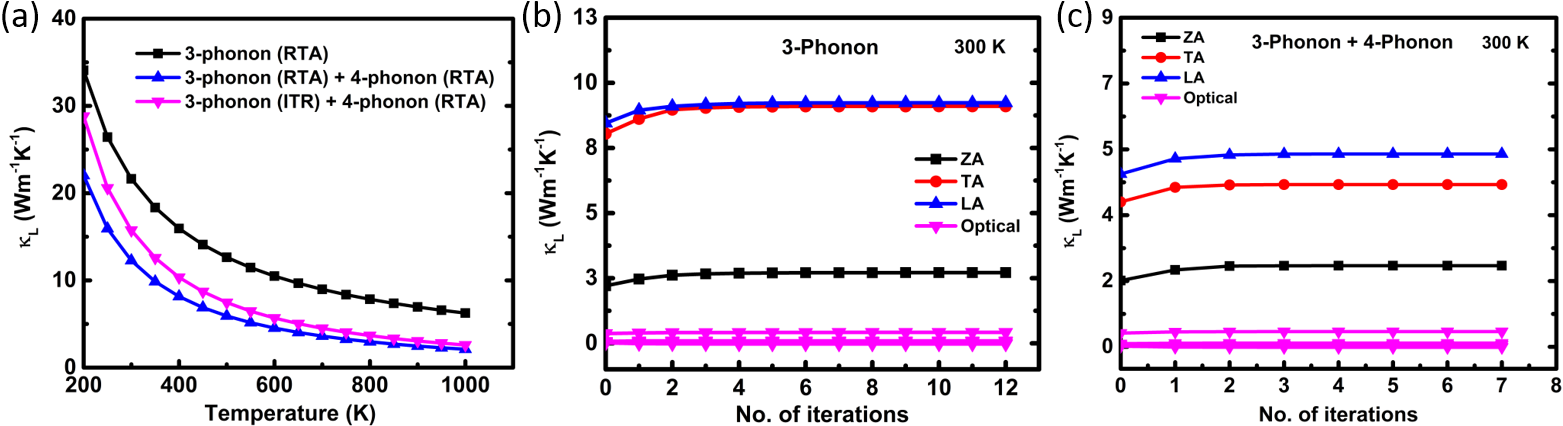}
	\caption{ (a) Variation in the $\kappa_{\text{L}}$ of ML-MoS$_{2}$ plotted as a function of temperature in the 8\% biaxially strained condition (BI$_{8}$). The plot includes results obtained with and without incorporating the 4-phonon scattering for both RTA and ITR. Contribution in $\kappa_{\text{L}}$ from the individual phonon modes of strained ML-MoS$_{2}$ (BI$_{8}$) at 300 K considering (b) 3-phonons only and (c) both 3-phonon and 4-phonon scattering rates.}
	\label{Fig.12}
\end{figure}

Finally, the calculated phonon scattering rates are incorporated to compute the lattice thermal conductivity ($\kappa_{\text{L}}$) of ML-MoS$_{2}$ in the biaxial 8\% strained condition (BI$_{8}$). The variation in $\kappa_{\text{L(3)}}$ and $\kappa_{\text{L(3+4)}}$, based on $\tau_{3}^{-1}$ and $\tau_{3}^{-1}$ + $\tau_{4}^{-1}$, respectively, as a function of temperature are presented in Fig. \ref{Fig.12}. The values of $\kappa_{\text{L(3)}}$ and $\kappa_{\text{L(3+4)}}$ at 300 K are found to be 21.6 and 12.2 Wm$^{-1}$K$^{-1}$, respectively. Compared to the $\approx$ 80\% reduction in $\kappa_{\text{L}}$ in the unstrained case, the inclusion of the 4-phonon scattering brings only $\approx$ 40\% reduction in the strained case. It is, therefore clear that, resulting from the low values of $\tau_{4}^{-1}$, the significance of 4-phonon scattering is much less in the strained case as compared to the unstrained one. From the plot of contribution of individual phonon modes to the total $\kappa_{\text{L}}$, it is seen that apart from the optical modes, the lowest contribution comes from the ZA mode (see Fig. \ref{Fig.12}(b)). The results remain same after the incorporation of the 4-phonon scattering as well, except the quantitative changes. The ZA phonons having the highest magnitudes of the Gr$\ddot{\text{u}}$neisen parameter ($\gamma$) resulting in the lowest values of $\kappa_{\text{L}}$ agrees well with the intuition. The accurate prediction of the role of ZA mode in thermal conduction considering only 3-phonon scattering proves the reduced importance of 4-phonon scattering with strain. To understand the obligation to go beyond the RTA solution, we have calculated the $\tau_{4}^{-1}$ corresponding to the Normal (N) and Umklapp (U) processes for the 8\% strained MoS$_{2}$. The variation in the $\tau_{4}^{-1}$ associated with the N and U processes as a function of the phonon frequency is shown in Fig. S9 and as a function of the wavevector along the k-path $\Gamma$-M is presented in Fig. S10 (see Supplementary Information). Although the percentage of the U processes has increased in the strained condition compared to the unstrained case, the N processes are still significant. Therefore, the ITR method has to be employed in the strained case as well for better accuracy. Similar to the unstrained case, solving the 4-phonon processes within the RTA level and inserting into the iterative scheme, the value of $\kappa_{\text{L(3+4)}}$ at 300 K is found to be 15.7 Wm$^{-1}$K$^{-1}$. Thereby, using both 3- and 4-phonon scattering we predict a $\approx$ 43\% reduction in $\kappa_{\text{L}}$ of ML-MoS$_{2}$ with the application of 8\% in-plane biaxial tensile strain. This is significantly less compared to the reduction of $\approx$ 82\% predicted using 3-phonon scattering only. In the absence of experimental validation for the thermal conductivity of ML-MoS$_{2}$ in the strained condition, results from the MD studies can be considered as a reliable benchmark. The prediction of 43\% reduction in $\kappa_{\text{L}}$ with 8\% biaxial strain, obtained when both the 3-phonon and 4-phonon scattering is considered, agrees well with the MD prediction of 60\% reduction with 12\% uniaxial strain \cite{ding2015manipulating}.

\begin{figure}[h!]
	\centering
	\includegraphics[scale=0.3]{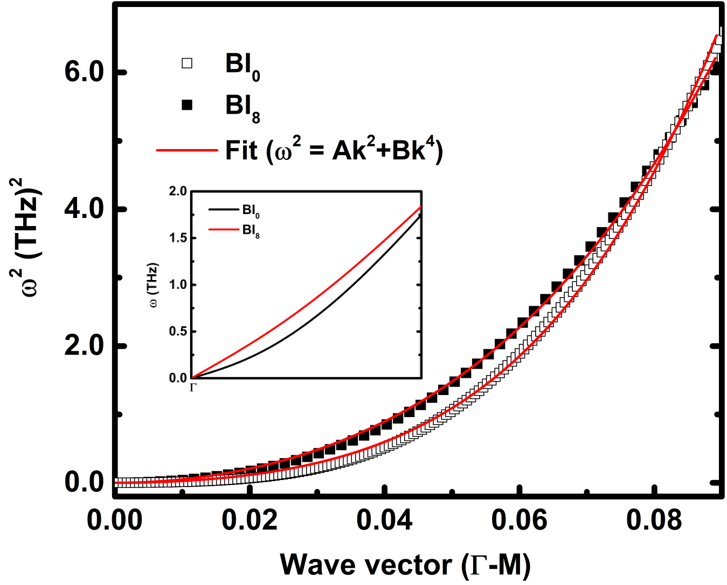}
	\caption{ Plot of $\omega^{2}$ as a function of the wave vector (k) around the $\Gamma$-point along the k-path $\Gamma$-M along with the fitting function $\omega^{2}$ = Ak$^{2}$ + Bk$^{4}$. Inset: ZA phonon mode dispersion close to the $\Gamma$-point of the BZ for the unstrained (BI$_{0}$) and the 8\% biaxially strained (BI$_{8}$) ML-MoS$_{2}$.}
	\label{Fig.13}
\end{figure}

Despite meeting the necessary conditions, such as the a-o frequency gap, reflection symmetry and strong anharmonicity, to have large 4-phonon scattering strength, the ML-MoS$_{2}$ in the strained condition (BI$_{8}$) is seen to have relatively less 4-phonon scattering rate. Among the three acoustic modes, the scattering rate ratio ($\tau_{4}^{-1}$/$\tau_{3}^{-1}$) is found to be the lowest for the ZA mode. This unusual behavior can be understood from the change in the ZA mode dispersion character with strain. The application of strain lowers the quadraticity of the ZA phonon mode and thereby, turns the ZA mode dispersion into a nearly linear one in the 8\% biaxially strained condition. To quantify the change in the quadraticity, we fitted the dispersion of the ZA mode near the BZ centre ($\Gamma$-point) along $\Gamma$-M with a model equation $\omega^{2}$ = Ak$^{2}$ + Bk$^{4}$, where $\omega$ is the frequency of the ZA mode at a wave vector k (see Fig. \ref{Fig.13}). The coefficients, A and B, are related to the linear and quadratic character of the dispersion, respectively. From the fitting, the values of A and B obtained in the unstrained condition are 260 and 70402, whereas in the strained condition the values become 505 and 35037. It is, therefore clear that, with strain application, the linear character increases and the quadratic character decreases in the dispersion of the ZA mode. The change in the dispersion from quadratic to linear results in a large reduction in the low-energy phonon population. Thereby, the phase space corresponding to the 4-phonon processes involving ZA phonons decrease. Similar phenomena have been observed in graphene as well [30]. It is, therefore, clear that merely having the a-o frequency gap, reflection symmetry and strong anharmonicity does not guarantee a large 4-phonon scattering strength. It also suggests a strong coupling between the dispersion characteristics of the ZA mode and the associated 4-phonon scattering rate of ML-MoS$_{2}$.

\section{Conclusions}
In summary, using first principle calculations we have demonstrated that the 4-phonon scattering rate is surprisingly high in ML-MoS$_{2}$. Owing to the quadratic dependency of $\tau_{4}^{-1}$ on temperature, the 4-phonon scattering dominates the overall phonon scattering processes at high temperatures. This unusually high $\tau_{4}^{-1}$ compared to the $\tau_{3}^{-1}$ is due to the restricted 3-phonon phase space resulting from the acoustic-optical frequency gap, strong anharmonicity and reflection symmetry in ML-MoS$_{2}$. The inclusion of the 4-phonon scattering significantly reduces the lattice thermal conductivity ($\kappa_{\text{L}}$) of ML-MoS$_{2}$. The error in the prediction of $\kappa_{\text{L}}$ with the experimental value reduces from $\approx$ 230\% in the calculation with $\tau_{3}^{-1}$ only to $\approx$ 30\% when both $\tau_{3}^{-1}$ and $\tau_{4}^{-1}$ are considered. Additionally, due to the significant presence of the Normal scattering processes in ML-MoS$_{2}$, the relaxation time approximation (RTA) results in an underestimation of $\kappa_{\text{L}}$. We, therefore, conjecture that an iterative solution for both 3- and 4-phonon processes will lead to the best accurate prediction of $\kappa_{\text{L}}$ for ML-MoS$_{2}$. We found that the $\kappa_{\text{L}}$ associated with the acoustic phonon modes in general, and the ZA mode in particular, is brutally overpredicted when only 3-phonon scattering is considered. Although the essential conditions to have large $\tau_{4}^{-1}$ is satisfied, the ML-MoS$_{2}$ in the strained condition is found to have a remarkably low strength of 4-phonon scattering. The significance of the quadratic dispersion of the ZA branch and therefore, the high population of the low-energy ZA phonons behind the strong 4-phonon scattering is revealed through systematic calculations. The results provide a critical revisit to the existing theories concerning the thermal transport properties of ML-MoS$_{2}$, and thus, any 2D material in general. Our work, therefore, significantly advances the thermal transport calculations towards accurate prediction of thermal conductivity of ML-MoS$_{2}$ and any other 2D TMDCs with analogous crystal structure.

\begin{acknowledgments}
The first-principles calculations have been performed using the supercomputing facility of IIT Kharagpur established under the National Supercomputing Mission (NSM), Government of India and supported by the Centre for Development of Advanced Computing (CDAC), Pune. AB acknowledges SERB POWER grant (SPG/2021/003874) and BRNS regular grant (BRNS/37098) for the financial assistance. SC acknowledges MHRD, India, for financial support.
\end{acknowledgments}
	
\bibliographystyle{achemso}
\bibliography{biblio}
	
\end{document}